\documentstyle[amsfonts,aps,twocolumn,epsfig]{revtex}
\begin{document}
\draft
\title{On a generalization of the logistic map}
\author{Val\'{e}rie Poulin\cite{vp}}
\address{Department of Mathematics and Statistics, Carleton University, Carleton,
Ontario, Canada K1S 5B6}
\author{Hugo Touchette\cite{ht}}
\address{Department of Physics and School of Computer Science, McGill University,
Montr\'{e}al, Qu\'{e}bec, Canada, H3A 2A7}
\date{\today}
\maketitle

\begin{abstract}
A family of non-conjugate chaotic maps generalizing the well-known logistic
function is defined, and some of its basic properties studied. A simple
formula for the Lyapunov exponents of all the maps contained in this family
is given based on the construction of conjugacies. Moreover, it is shown
that, despite the dissimilarity of their polynomial expressions, all the
maps possess the same invariant density. Other algebraic properties of the
family, which shows some relationship with the set of Tschebysheff
polynomials, are also investigated.
\end{abstract}


\hrule\medskip

\noindent {\bf Simple iterated maps, such as the Baker map and the logistic
map, are the subject of constant fascination. Part of the interest for these
systems is linked to the fact that they provide an easy and pedagogical way
to understand how complex and chaotic behavior can arise from simple
dynamical models. Even more remarkable, yet, is the fact that studies of
low-dimensional maps have proven to be fruitful in understanding the basic
mechanisms responsible for the appearance of chaos in a large class of
dynamical systems (e.g., differential flows, high-dimensional maps). One
paradigm example of such mechanisms is the so-called period-doubling
cascades of fixed points, encountered qualitatively in many physical systems
of interest. In this paper, we enlarge the set of maps known to be chaotic
by presenting a generalization of the logistic map. The generalization, more
precisely, enables us to construct an infinite number of one-dimensional
maps which are chaotic in the sense that they all have positive Lyapunov
exponents, and possess at least one orbit that is not asymptotically
periodic. }

\medskip\hrule

\section{Introduction}

In the theory of nonlinear systems, the logistic map 
\begin{equation}
x_{n+1}=f_r(x_n)=rx_n(1-x_n),  \label{log}
\end{equation}
with $0\leq r\leq 4$, $x_n\in [0,1]$, and $n=0,1,2,\ldots $, is well-known
to provide one of the simplest example of what is referred to as a {\it %
chaotic} system. That is, a system which, for some range of its parameters,
possesses at least one bounded orbit $\{x_0,x_1,\ldots \}$ such that (i) $%
\{x_0,x_1,\ldots \}$ is not asymptotically periodic and (ii) the Lyapunov
exponent, defined in the context of Eq.(\ref{log}) by the usual relation 
\begin{equation}
\lambda (x_0)=\lim_{N\rightarrow \infty }\sum_{i=0}^{N-1}\ln \left| \partial
_xf_r(x_i)\right| ,  \label{lyap}
\end{equation}
is greater than zero \cite{yorke,note1}. For the logistic map at $r=4$,
specifically, $\lambda (x_0)$ is positive and equals $\ln 2$ for almost all $%
x_0$. This value can be easily calculated by noting that the logistic map is
conjugate to the tent map or by having recourse to the fact that the
dynamical system defined by Eq.(\ref{log}) is ergodic for this particular
value of $r$ \cite{robinson,beck}. In this latter case, one is justified to
calculate analytically the Lyapunov exponent as an ensemble average 
\begin{equation}
\lambda =\int_0^1\rho (x)\ln \left| \partial _xf_r(x)\right| dx,
\end{equation}
the function 
\begin{equation}
\rho (x)=\frac 1{\pi \sqrt{x(1-x)}},\qquad x\in [0,1]  \label{inv1}
\end{equation}
being the invariant density of the map satisfying $\rho (A)=\rho (f^{-1}(A))$%
, where $f^{-1}(A)$ is the preimage of an arbitrary subset $A$ of the unit
interval \cite{beck,ruelle}.

In this paper, we shall focus on another property of the logistic map at $r=4
$, namely the existence of the following closed-form solution 
\begin{equation}
x_n=\sin ^2(2^n\pi \theta _0),  \label{clo}
\end{equation}
where $\theta _0=\pi ^{-1}\arcsin (\sqrt{x_0})$. The importance of this
formula relies evidently on the fact that $x_n$ can be evaluated directly,
for any initial point $x_0$, without actually computing the intermediate
values $x_1,x_2,\ldots ,x_{n-1}$. In the remaining of this work, we follow
the steps used in the derivation of the above formula to construct similar
expressions involving general terms of the form $\sin ^2(N^n\theta )$. In
doing so, we shall see that a whole new family of polynomial maps can be
defined which, essentially, generalize the logistic map by preserving the
density Eq.(\ref{inv1}), and whose Lyapunov exponents can be calculated
easily. Other properties of these newly defined dynamical maps are also
studied. In particular, we shall point out some similarities between the
maps and the Tschebysheff polynomials, and prove, finally, that the family
of maps as a whole is closed under the composition of functions.

\section{Definition of the logistic family}

The explicit solution of the logistic map at the particular value $r=4$ can
be derived by substituting in $f_4(x_n)=4x_n(1-x_n)$ the change of
coordinates $x_n=\sin ^2(\pi \theta _n)$, valid for $x_n\in [0,1]$, and by
using the trigonometric identity 
\begin{equation}
4\sin ^2\theta (1-\sin ^2\theta )=\sin ^2(2\theta )
\end{equation}
in order to obtain $\sin ^2(\pi \theta _{n+1})=\sin ^2(2\pi \theta _n)$.
This is equivalent to the map $\theta _{n+1}=2\theta _n%
%TCIMACRO{\func{mod}}
%BeginExpansion
\mathop{\rm mod}%
%EndExpansion
1$, which has the explicit solution 
\begin{equation}
\theta _n=2^n\theta _0%
%TCIMACRO{\func{mod} }
%BeginExpansion
\mathop{\rm mod}%
%EndExpansion
1.  \label{smap}
\end{equation}
Hence, the complete solution in terms of the coordinate $x$ must correspond
to Eq.(\ref{clo}).

Following these steps, one can imagine to construct a map having the
solution $x_n=\sin (3^n\pi \theta _0)$ by expressing $\sin ^2(3\theta )$ in
terms of $\sin ^2(\theta )$ with the identity 
\begin{equation}
\sin ^2(3\theta )=16\sin ^6\theta -24\sin ^4\theta +9\sin ^2\theta .
\end{equation}
In this case, one gets $g(x)=16x^3-24x^2+9x$ as a possible dynamical map on $%
[0,1]$ having the required solution. More generally, one can use the
recurrence formula 
\begin{equation}
\sin (N\theta )=2\cos \theta \sin [(N-1)\theta ]-\sin [(N-2)\theta ]
\label{rec}
\end{equation}
to define a whole set of maps which express $\sin ^2(N\theta )$ in terms of $%
\sin ^2\theta $. This set, the {\it sine functions set}, is defined
specifically as 
\begin{equation}
{\cal S}=\{SN(x):N=1,2,3,\ldots \},
\end{equation}
where $SN(x)=[s_N(\sqrt{x})]^2$, and 
\begin{eqnarray}
s_1(x) &=&x  \nonumber \\
s_2(x) &=&2x\sqrt{1-x^2}  \nonumber \\
&\vdots &  \nonumber \\
s_N(x) &=&2\sqrt{1-x^2}s_{N-1}(x)-s_{N-2}(x).
\end{eqnarray}
Note that the intermediate functions $s_N(x)$ are expressed as such in order
to verify Eq.(\ref{rec}) with the variable change $x=\sin \theta $.

\begin{center}
\bigskip

\begin{tabular}{c}
\hline\hline
$SN(x)$ \\ \hline
\multicolumn{1}{l}{$S1(x)=x$} \\ 
\multicolumn{1}{l}{$S2(x)=-4x^2+4x$} \\ 
\multicolumn{1}{l}{$S3(x)=16x^3-24x^2+9x$} \\ 
\multicolumn{1}{l}{$S4(x)=-64x^4+128x^3-80x^2+16x$} \\ 
\multicolumn{1}{l}{$S5(x)=256x^5-640x^4+560x^3-200x^2+25x$} \\ \hline\hline
\end{tabular}

\medskip

Table 1. First five members of ${\cal S}$.
\end{center}

\newpage

Using these definitions, we calculate the first five functions of ${\cal S}$
listed in table 1. Obviously, by construction of ${\cal S}$, $S2(x)$ is the
logistic equation itself with parameter $r=4$. From a more general
perspective, it can also be seen that the maps $SN(x)$ are degree $N$
polynomials whose leading coefficient, i.e., the coefficient of the highest
degree term, is equal to $4^{N-1}$ in absolute value. These two results,
satisfied by any function $SN(x)$, is proven more formally in ref.\cite
{poulin}. Note that the latter property allows us to extend the similarity
with the logistic map by parameterizing the functions of ${\cal S}$ in the
following manner 
\begin{equation}
SN_r(x)=r\frac{SN(x)}{4^{N-1}},
\end{equation}
with $0\leq r\leq 4^{N-1}$. We call the set of functions $\{SN_r(x):0\leq
r\leq 4^{N-1}\},$ the {\em family of } $SN(x)$, which can be characterized
numerically by bifurcation diagrams and Lyapunov spectrums such as the ones
shown in figure%
%TCIMACRO{\TeXButton{~}{~}}
%BeginExpansion
~%
%EndExpansion
1.

\begin{figure}[t]
\centering
\epsfig{file=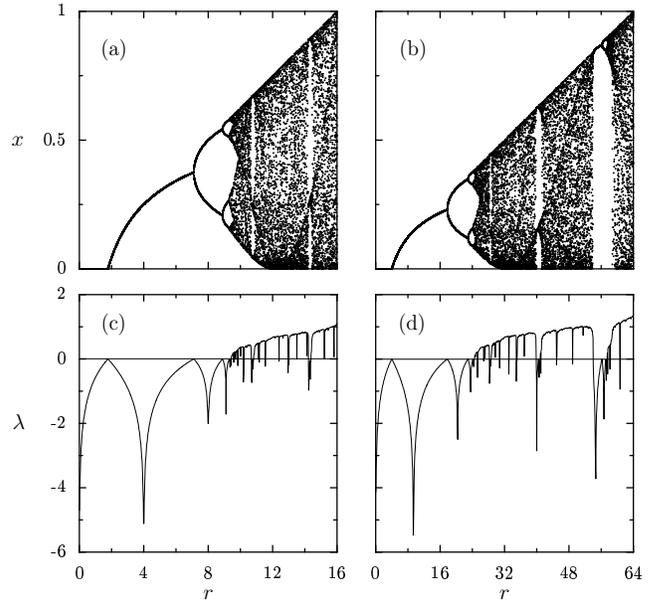,width=3.4in}

\caption{(a)-(b) Bifurcation diagrams of $S3(x)$ and $S4(x)$. (c)-(d)
Corresponding Lyapunov spectrums.}
\end{figure}

\section{Particularities of ${\cal S}$}

Many of the interesting properties of the logistic map at $r=4$ can be
investigated more intuitively by making explicit the fact that Eq.(\ref{smap}%
) is equivalent to a shift map $S$ on the binary expression of $\theta _0$ 
\cite{yorke}. Indeed, if we express $\theta _0$ as a binary number 
\begin{equation}
\theta _0=0.b_0b_1b_2\ldots =\sum_{i=0}^\infty \frac{b_i}{2^{i+1}},\qquad
b_i\in \{0,1\},
\end{equation}
then applying Eq.(\ref{smap}) to $\theta _0$ is equivalent to shifting all
the bits of $\theta _0$ to the left and dropping the integer part. In other
words, 
\begin{eqnarray}
\theta _n &=&0.b_nb_{n+1}b_{n+2}\ldots  \nonumber \\
&=&S^n(0.b_0b_1b_2\ldots ),
\end{eqnarray}
where $S^n=S\circ S^{n-1}$ for $n>1$, and $S^1=S$. Not surprisingly, the
same is true for the maps $SN(x)$, since Eq.(\ref{smap}) was the guideline
in defining the family ${\cal S}$. However, in the case of $SN(x)$, the
shift map to consider takes effect on $\theta _0$ written in base $N$. This
follows from the following result which generalizes effectively the solution
of Eqs.(\ref{clo}) and (\ref{smap}).

{\it Theorem 1.} Let $\{x_0,x_1,x_2,\ldots \}$ be the orbit of $x_0$ under $%
SN(x)$. If we write $x_n=\sin ^2(\pi \theta _n)$, then we have that 
\begin{equation}
\theta _{n+1}=N^n\theta _0%
%TCIMACRO{\func{mod}}
%BeginExpansion
\mathop{\rm mod}%
%EndExpansion
1,  \label{smapm}
\end{equation}
where, as usual, $\theta _0=\pi ^{-1}\arcsin (\sqrt{x_0})$.

We omit the proof of this theorem as it follows directly from the next lemma.

{\it Lemma 1}. Consider $SN(x)$ as defined previously. We have that 
\begin{equation}
SN(\sin ^2\theta )=\sin ^2(N\theta ).  \label{sin1}
\end{equation}

{\it Proof:} The result is obvious for $N=1$ and $N=2$. Suppose Eq.(\ref
{sin1}) true for $N-1$ and $N-2$, that is to say 
\begin{eqnarray*}
s_{N-1}^2(\sin \theta ) &=&S[N-1](\sin ^2\theta ) \\
&=&\sin ^2[(N-1)\theta ],
\end{eqnarray*}
and $s_{N-2}(\sin \theta )=\sin [(N-2)\theta ]$. Then, for $\theta \in {\Bbb %
R},$%
\begin{eqnarray*}
SN(\sin ^2\theta ) &=&s_N^2(\sin \theta ) \\
&=&[2\sqrt{1-\sin ^2\theta }s_{N-1}(\sin \theta )-s_{N-2}(\sin \theta )]^2 \\
&=&[2\cos \theta \sin ((N-1)\theta )-\sin ((N-2)\theta )]^2 \\
&=&\sin ^2(N\theta ),
\end{eqnarray*}
where we have used the identity (\ref{rec}).\hfill $\Box $

Note that we could have proceeded to a similar generalization of the
logistic map using cosine functions instead of sine functions, while
preserving its shift property. One possible way of achieving this is to
define the {\it cosine functions set} 
\begin{equation}
{\cal C}=\{CN(x):N=1,2,3,\ldots \},
\end{equation}
where $CN(x)=[c_N(\sqrt{x})]^2$, and 
\begin{eqnarray}
c_1(x) &=&x  \nonumber \\
c_2(x) &=&2x^2-1  \nonumber \\
&\vdots &  \nonumber \\
c_N(x) &=&2xc_{N-1}(x)-c_{N-2}(x).  \label{rec1}
\end{eqnarray}

Contrary to the $s_N(x)$'s, the functions $c_N(x)$ have the interesting
property that they are polynomials of degree $N$. In fact, the set $%
\{c_N(x):N\in {\Bbb N}\}$ coincides with the set of Tschebysheff polynomials
on the unit interval \cite{crc}, the latter set satisfying the exact same
recurrence formula as Eq.(\ref{rec1}). We thus have that $\{c_N(x)\}$ must
constitute a set of orthogonal polynomials, i.e., 
\begin{equation}
\int_0^1c_N(x)c_{N^{\prime }}(x)\ dx=\delta _{N,N^{\prime }},
\end{equation}
for all integers $N$ and $N^{\prime }$, where $\delta _{i,j}$ is the
delta-Kronecker function. This fact can be further proved using the property 
$c_N(\cos \theta )=\cos (N\theta )$, well-known to be satisfied by the
Tschebysheff functions. Note that $\{s_N(x)\}$ is also a set of orthogonal
functions; its members satisfy indeed the relation $s_N(\sin \theta )=\sin
(N\theta )$. In the remaining of this work, we shall restrain our study to
the set ${\cal S}$, since the maps $SN(x)$ are directly related to $CN(x)$
by the expression 
\begin{equation}
SN(x)=\left\{ 
\begin{array}{lll}
CN(x), &  & \text{for }N\text{ odd} \\ 
1-CN(x), &  & \text{for }N\text{ even.}
\end{array}
\right.
\end{equation}
Hence, as far as their dynamics are concerned, the functions $SN(x)$ and $%
CN(x)$ are totally equivalent.

\section{Conjugacies}

The analysis of the chaoticity properties of a map $f$ is greatly simplified
by studying conjugate maps of $f$ which are obtained by applying a global
change of variables. Recall that two maps $f:I\rightarrow I$ and $%
g:J\rightarrow J$ are {\it conjugate} if there exists a homeomorphism, i.e.,
a bijective and continuous map $H:I\rightarrow J$ such that $H\circ f=g\circ
H$. The function $H$ is called a {\it conjugacy}. In the context of ${\cal S}
$, a possible conjugate function of $SN(x)$ can be constructed as follows.
Let $TN(x)$ be a piecewise linear function (a generalized tent map) defined
on subintervals $[k/N,(k+1)/N]$ of $[0,1]$ by setting 
\begin{equation}
TN(x)=\left\{ 
\begin{array}{ll}
Nx-k, & \text{for }k\text{ even} \\ 
-Nx+k+1, & \text{for }k\text{ odd,}
\end{array}
\right. 
\end{equation}
with $k=0,1,\ldots ,N-1$.

{\it Theorem 2}. $SN(x)$ is conjugate to $TN(x)$ with conjugacy $H(x)=\sin
^2(\pi x/2)$.

{\it Proof}: First note that $H(x)$ is both continuous and bijective on the
interval $[0,1]$. Now, on the first hand we have that $SN(H(x))=SN(\sin
^2(\pi x/2))=\sin ^2(N\pi x/2)$. On the other hand, 
\begin{eqnarray}
H(TN(x)) &=&\left\{ 
\begin{array}{ll}
\sin ^2(N\pi x/2-k\pi /2), & \text{for }k\text{ even} \\ 
\sin ^2(-N\pi x/2+(k+1)\pi /2), & \text{for }k\text{ odd}
\end{array}
\right.  \nonumber \\
&=&\left( \pm \sin (N\pi x/2)\right) ^2  \nonumber \\
&=&\sin ^2(N\pi x/2).
\end{eqnarray}
Thus, we have proved that $SN\circ H=H\circ TN$ for all integers $N$.\hfill $%
\Box $

Figure 2 depicts the graphs of $SN(x)$ and the corresponding $TN(x)$ for $%
N=3,4$. ($T2(x)$ is only the tent map since, as we mentioned, $S2(x)$ is the
logistic map.) In each case, note that $SN(x)$ is a ``smooth'' version of $%
TN(x)$.

{\it Remark}. The functions $CN(x)$ are pairwise conjugate to the functions $%
TN(x)$ with conjugacy map $H(x)=\cos ^2(\pi x/2)$. Evidently, since the
conjugacy is an equivalence relation, this has the consequence that any $%
CN(x)$ is conjugate to $SN(x)$.

\begin{figure}[b]
\centering
\epsfig{file=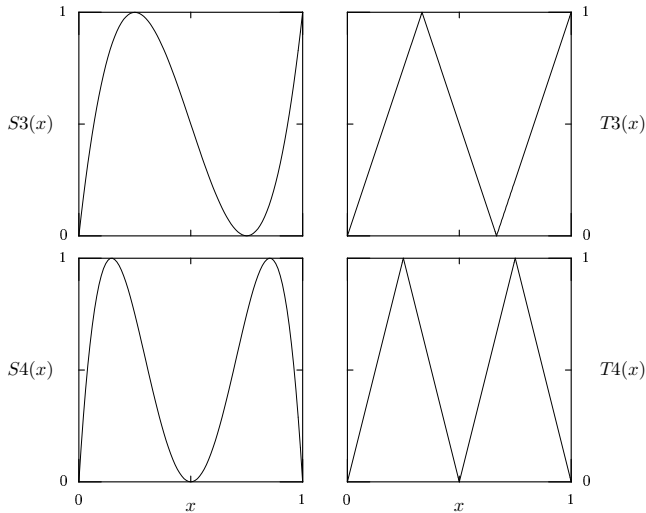,width=3.4in}

\caption{Graphs of the generalized tent maps $T3(x)$ and $T4(x)$ which are 
conjugate to $S3(x)$ and $S4(x)$.}
\end{figure}

\section{Ergodic properties of ${\cal S}$}

In this section, we evaluate the invariant densities of the maps $SN(x)$ and
their Lyapunov exponents based on the conjugates found in the previous
section. As one would expect from the similarity of the graphs of $T3(x)$
and $T4(x)$ in figure 2, the functions $TN(x)$ should exhibit a unique and
constant invariant density 
\begin{equation}
\rho _{TN}(x)=1,\qquad x\in [0,1],
\end{equation}
just as it is the case for the tent map. This can be proven using Adler and
Bowen's results on Markov transformations \cite{mane}, (see also \cite
{poulin}). At this point, the invariant density of the functions $SN(x)$ can
be found using the transformation formula 
\begin{equation}
\rho _{SN}(x)=\rho _{TN}(y)\left| \frac{dy}{dx}\right| ,
\end{equation}
where $y=\frac 2\pi \arcsin (\sqrt{x}),$ so as to obtain 
\begin{equation}
\rho _{SN}(x)=\frac 1{\pi \sqrt{x(1-x)}}.
\end{equation}
The unicity of $\rho _{SN}(x)$ is assured by the fact that $\rho _{TN}(x)$
is also unique, which means that both functions $SN(x)$ and $TN(x)$ must be
ergodic \cite{robinson,beck,mane}.

Now, to evaluate the Lyapunov exponent of $SN(x)$ we may use the fact that $%
\left| \partial _xTN(x)\right| =N$ for almost all $x\in [0,1]$ to infer that 
$\lambda (x_0)=\ln N$ almost everywhere in the case of $TN(x)$. Accordingly,
since Lyapunov exponents are invariant under smooth and differentiable
coordinate transformations \cite{mccauley}, we have the following theorem.
(A more extensive proof of this result, which takes care of the pathological
points where $\partial _xTN(x)$ is not defined, is contained in ref.\cite
{poulin}.)

{\it Theorem 3.} The Lyapunov exponent of $SN(x)$ is $\ln N$ almost
everywhere (with respect to the invariant measure $\rho _{SN}(x)$).

The above theorem shows that the members of ${\cal S}$ are non-conjugate to
each other simply because they possess different Lyapunov exponents. It also
shows that ${\cal S}\backslash \{S1\}$, and consequently the set of
Tschebysheff polynomials, are sets of {\it chaotic} maps. Indeed, $\lambda
=\ln N>0$ for $N>1$, and by using the shift property of $SN(x)$ we can
choose $x_0=\sin ^2(\pi \theta _0)$, with $\theta _0$ irrational, to build
an orbit that is not asymptotically periodic. Another way to convince
ourselves that all the polynomials in ${\cal S}$ have chaotic orbits is to
use the celebrated result ``period-3 implies chaos'' \cite{yorke}, and find
an initial point $x_0$ of period 3 for each $SN(x)$. For instance, for a $N>1
$ let $x_0=\sin ^2(\pi \theta _0)$ where 
\begin{equation}
\theta _0=\frac 1{N^3-1}=0.001001\ldots \qquad \text{(in base }N\text{).}
\end{equation}
Again, using the shift map property, we must have 
\begin{eqnarray}
x_0 &=&\sin ^2(\pi \cdot 0.001001\ldots )  \nonumber \\
x_1 &=&\sin ^2(\pi \cdot 0.010010\ldots )  \nonumber \\
x_2 &=&\sin ^2(\pi \cdot 0.100100\ldots )  \nonumber \\
x_3 &=&\sin ^2(\pi \cdot 0.001001\ldots )=x_0.
\end{eqnarray}
We thus extended the chaoticity properties of the logistic map to an
infinite family of polynomials.

\section{Algebraic properties}

To complete the study of the properties of ${\cal S}$, we now deduce that it
is an abelian monoid with respect to the composition of functions ($\circ $%
). A monoid, precisely, is a non-empty set $M$ together with a binary
associative operation, say $*$, such that $x*y\in M$ for $x,y\in M$. There
must also be an element $e\in M$, called the identity element, for which $%
x*e=e*x=x$ for all $x\in M$. Moreover, a monoid is called abelian if the
binary operation is commutative \cite{jacobson}. In ${\cal S}$, the identity
element is $S1(x)=x$. Also, the composition of function is clearly
associative. Now, to prove that ${\cal S}$ is indeed abelian monoid, we
verify that it is closed under composition and that this composition is
commutative, a condition that is not verified in the case of composition of
general functions. However, before we do so, we present next a new
expression of $SN(x)$ on the unit interval.

{\it Lemma 2.} For all $SN(x)$ and $x\in [0,1]$, 
\[
SN(x)=\sin ^2(N\arcsin \sqrt{x}). 
\]

{\it Proof:} Let $x\in [0,1]$. There exists a $\theta \in [0,\pi /2]$ such
that $x=\sin ^2\theta $, and thus $\theta =\arcsin \sqrt{x}$. Now, from
Lemma 1 we have $SN(x)=SN(\sin ^2\theta )=\sin ^2(N\theta )=\sin ^2(N\arcsin 
\sqrt{x})$.\hfill $\Box $

{\it Theorem 4.} (Monoid property) Let $N_1$ and $N_2$ be any positive
integers. We have that 
\begin{equation}
SN_1\circ SN_2=SN_2\circ SN_1=S[N_1N_2].
\end{equation}

{\it Proof:} For $N_1$ and $N_2$ given, consider $SN_1(x)$ and $SN_2(x)$.
Then, for any $x\in [0,1]$, we obtain from Lemma%
%TCIMACRO{\TeXButton{~}{~}}
%BeginExpansion
~%
%EndExpansion
2 
\begin{eqnarray}
SN_1(SN_2(x)) &=&\sin ^2[N_1\arcsin (\sqrt{SN_2(x)})]  \nonumber \\
&=&\sin ^2[N_1\arcsin (\sin (N_2\arcsin \sqrt{x}))]  \nonumber \\
&=&\sin ^2(N_1N_2\arcsin \sqrt{x})  \nonumber \\
&=&S[N_1N_2](x).
\end{eqnarray}
Obviously, $S[N_1N_2](x)=S[N_2N_1](x)$, so the composition is
commutative.\hfill $\Box $

As a direct consequence of the monoid property, $k$-periodic points of a
certain polynomial $SN(x)$ can be looked at as fixed points of the function $%
SM(x)$ where $M=N^k$. Furthermore, a polynomial $SN(x)$ of very high degree
can be computed easily by decomposing its expression using lower degree
polynomial of the family ${\cal S}$. Explicitly, consider $SN(x)\in {\cal S}$%
. We say that $SN(x)$ is a {\it prime element} of ${\cal S}$ if $N$ is a
prime number. Using this definition, we have as a result of Theorem 4 and
the Fundamental Theorem of Arithmetic that any polynomial $SN(x)$ must
possess a unique decomposition in prime elements of ${\cal S}$.

\section{Final remarks}

To conclude, note that our study of the sine functions, written in the form $%
SN(x)=\sin ^2(N\arcsin \sqrt{x})$, have been restricted to positive integers 
$N$. In a similar manner, it could be interesting to investigate functions
of the type $S\alpha (x)=\sin ^2(\alpha \arcsin \sqrt{x})$ with $\alpha $
real. One observation about this extra generalization is that, as for $S1(x)$%
, the function $S\alpha (x)$ does not exhibit chaotic properties for $0\leq
\alpha \leq 1$. The function $S\frac 12(x)=\sin ^2(\frac 12\arcsin \sqrt{x})$%
, for example, is conjugate to $g(x)=x/2$, and has all of its orbits
attracted to $x=0$. Yet, this is not surprising since the Lyapunov exponent
of this map must be $\ln (1/2)<0$. This brings us to conjecture that $%
S\alpha (x)$ must admit chaotic behavior if and only if $\left| \alpha
\right| >1$, considering that the Lyapunov exponents of $S\alpha (x)$ should
be $\ln \alpha $. A complete proof of this result, however, cannot be given
here using the same symbolic dynamic approach used for $SN(x)$, for the
simple reason that the expression of a point in ``base $N$'' makes sense
only if $N$ is an integer greater than $1$.

\section*{Aknowledgements}

V.P. would like to thank A. Mingarelli for helpful discussions. This work
was supported in part by the National Sciences and Engineering Research
Council of Canada (NSERC) through the ES A Scolarship program.


\begin{references}
\bibitem[{{*}}]{vp}  Electronic address: vpoulin@omnisig.com

\bibitem[{\dagger }]{ht}  Electronic address: htouc@cs.mcgill.ca

\bibitem{yorke}  K.T. Alligood, T.D. Sauer, J.A. Yorke, {\it Chaos: An
Introduction to Dynamical Systems} (Springer Verlag, New York, 1997).

\bibitem{note1}  This definition of chaos is not necessarily equivalent to
the more widely held definition based on topological transitivity and
sensitivity to initial conditions (see, e.g., refs.\cite{devaney,robinson}).
However, the conditions of chaoticity given in the text, which are used
throughout this work, are more easy to verify in practice, and nevertheless
capture most of the essense of what is understood as chaos.

\bibitem{devaney}  R.L. Devaney, {\it A First Course in Chaotic Dynamical
Systems: Theory and Experiment} (Addison-Wesley, Reading, 1992).

\bibitem{robinson}  C. Robinson, {\it Dynamical Systems: Stability, Symbolic
Dynamics, and Chaos}, 2nd ed (CRC Press, New York, 1999).

\bibitem{beck}  C. Beck, F. Schl\"{o}gl, {\it Thermodynamics of Chaotic
Systems} (Cambridge Univ. Press, Cambridge, 1993).

\bibitem{ruelle}  J.-P. Eckmann, D. Ruelle, {\it Rev. Mod. Phys. }{\bf 57},
617 (1985).

\bibitem{poulin}  V. Poulin, M.Sc. Thesis, Department of Mathematics and
Statistics, Carleton University, 1999. (unpublished)

\bibitem{crc}  D. Zwillinger (ed.), {\it Standard Mathematical Tables and
Formulae} (CRC Press, New York, 1996), 30th ed., p.489.

\bibitem{mane}  R. Ma\~{n}\'{e}, {\it Ergodic Theory and Differentiable
Dynamics} (Springer-Verlag, 1983).

\bibitem{mccauley}  J.L. McCauley, {\it Chaos, Dynamics, and Fractals}
(Cambridge Univ. Press, Cambridge, 1993).

\bibitem{jacobson}  N. Jacobson, {\it Basic Algebra I} (W.H. Freeman and
Company, New York, 1985).
\end{references}
\end{document}